\documentclass[aps,preprint]{revtex4}%
\usepackage{amsfonts}
\usepackage{amsmath}
\usepackage{amssymb}
\usepackage{graphicx}
\usepackage{pdfpages}
\usepackage{mathrsfs}
\usepackage{float}
\usepackage{mathrsfs}
\usepackage{color}
\usepackage{xcolor}
\usepackage{caption}
\usepackage{appendix}
\usepackage{lineno,hyperref}
\usepackage{changes}%
\hypersetup{colorlinks =true, allcolors = blue}
\providecommand{\U}[1]{\protect\rule{.1in}{.1in}}

\begin{document}
\begin{center}
{\Large Geodesics structure and deflection angle of electrically charged black holes in gravity with a background
Kalb-Ramond field}

Ahmad Al-Badawi 

Department of Physics, Al-Hussein Bin Talal University, P. O. Box: 20,
71111, Ma'an, Jordan
\bigskip E-mail: ahmadbadawi@ahu.edu.jo\\
Sanjar Shaymatov

Institute for Theoretical Physics and Cosmology, Zhejiang University of Technology, Hangzhou 310023, China\\
Institute of Fundamental and Applied Research, National Research University TIIAME, Kori Niyoziy 39, Tashkent 100000, Uzbekistan\\
University of Tashkent for Applied Sciences, Str. Gavhar 1, Tashkent 100149, Uzbekistan\\
Western Caspian University, Baku AZ1001, Azerbaijan
\bigskip E-mail: sanjar@astrin.uz\\
\.{I}zzet Sakall{\i}

Physics Department, Eastern Mediterranean University, Famagusta 99628, North Cyprus via Mersin 10, Turkey
\bigskip E-mail:izzet.sakalli@emu.edu.tr\\

{\Large Abstract}\end{center}
This article investigates the geodesic structure and deflection angle of charged black holes in the presence of a nonzero vacuum expectation value background of the Kalb-Ramond field. Topics explored include null and timelike geodesics, energy extraction by collisions, and the motion of charged particles. The photon sphere radius is calculated and plotted to examine the effects of both the black hole charge ($Q$) and the Lorentz-violating parameter ($b$) on null geodesics. The effective potential for timelike geodesics is analyzed, and second-order analytical orbits are derived. We further show that the combined effects of Lorentz-violating parameter and electric charge can mimic a Kerr black hole spin parameter up to its maximum values, i.e., $a/M \sim 1$ thus suggesting that the current precision of measurements of highly spinning black hole candidates may not rule out the effect of Lorentz-violating parameter. The center of mass energy of colliding particles is also considered, demonstrating a decrease with increasing Lorentz-violating parameter. Circular orbiting particles of charged particles are discovered, with the minimum radius for a stable circular orbit decreasing as both $b$ and $Q$ increase. Results show that this circular orbit is particularly sensitive to changes in the Lorentz-violating parameter. Additionally, a timelike particle trajectory is demonstrated as a consequence of the combined effects of parameters $b$ and $Q$. Finally, the light deflection angle is analyzed using the weak field limit approach to determine the Lorentz-breaking effect, employing the Gauss-Bonnet theorem for computation. Findings are visualized with appropriate plots and thoroughly discussed.

\section{Introduction}

Lorentz symmetry, a key notion in modern physics, states that all inertial reference frames obey the same physical laws. There have been several experiments and observations that have shown Lorentz symmetry to be a fundamental symmetry of nature. However, some theories have discovered that Lorentz symmetry can be broken at some energy scale. These include string theory \cite{ls1}, loop quantum gravity \cite{ls2}, Horava-Lifshitz gravity \cite{ls3}, non-commutative field theory \cite{ls4}, massive gravity \cite{ls6} and others \cite{ls7,ls8,ls9}. Studying Lorentz symmetry breaking (LSB) provides valuable insight into the nature of spacetime and fundamental physics principles. LSB might occur explicitly or spontaneously. The explicit LSB happens when the Lagrange density is not Lorentz invariant, which means that physical rules take distinct forms in different reference frames. However, spontaneous LSB happens when a physical system's ground state lacks Lorentz symmetry, while its Lagrange density is Lorentz invariant. The Standard-Model Extension \cite{ls10} provides a general framework for researching spontaneous LSBs. The simplest field theories presented within this paradigm are called bumblebee models \cite{ls11,ls12,isLessa:2023yvw,isOvgun:2018xys,isMangut:2023oxa,isKanzi:2022vhp,isCasana:2017jkc,isGullu:2020qzu,isLiu:2022dcn,isKanzi:2021cbg}. Vacuum expectation values (VEVs) are achieved in bumblebee models by using vector fields called bumblebee fields. In the presence of a nonzero VEV, particles are not invariant to Lorentz directions locally.\\ 
On the other hand, the Kalb-Ramond (KR) field \cite{kr13}, which may be described as a self-interacting second-rank antisymmetric tensor, modifies the Einstein action. The KR modification is connected to heterotic string theory \cite{kr14} and can be understood as closed string excitation. Because of the non-minimal coupling of the tensor field with the Ricci scalar, the Lorentz symmetry may be violated \cite{kr15}. Lessa et al. reported a precise static and spherically symmetric solution using the VEV backdrop of the KR field \cite{kr16}. Following that, the motion of massive and massless particles in the vicinity of this static spherical KR BH was investigated in reference \cite{kr17}. Ref. \cite{kr18} investigates the gravitational deflection of light and shadow cast by revolving KR BHs. Furthermore, the effects of the VEV backdrop on Bianchi type-I cosmology were investigated \cite{kr19} and the fermionic greybody factors and quasinormal modes of black holes (BHs)
in KR gravity were studied in \cite{aa1}.\\ Recently, an exact solution for static and spherically symmetric BHs in the setting of this Lorentz-violating gravity theory was reported in \cite{st1}.  They also investigate the physical ramifications of Lorentz violation by studying the thermodynamic features of these solutions and assessing their impact on several classical gravitational experiments in the Solar System. Later, \cite{RN1} presented electrically charged BHs in the absence and existence of a cosmological constant in gravity theory, with Lorentz violation caused by a background KR field. Moreover, they investigate the corresponding thermodynamic properties and shadow in this charged BH or Reissner-Nordstr\"{o}m-like (RN-like) BH \cite{RN1}. \textcolor{blue}{The study found that the shadow radius is highly sensitive to the Lorentz-violating parameter $b$, and decreases as $b$ increases}. It is to be emphasized that the KR modification as mentioned above is related to heterotic string theory. It does however differ from the other charged black hole solutions \cite{Shaymatov23EPJP,Shaymatov22a}.\\
It should be noted that, from the astrophysical point of view, it is increasingly important to gain a deeper understanding of the nature of the spacetime geometry and the existing fields that can significantly alter the geodesic structure of particles and photons and thus can influence observable properties including the innermost stable circular orbits (ISCO), the size of photon sphere/shadow size, etc which play a crucial role to help us understand qualitatively not only the background spacetime geometry but also the existing fields around astrophysical BHs. It does therefore have value to explore remarkable insights of the BH solution, which can provide striking differences from their mimickers in various theories of gravity in the regions close to the horizon. There are investigations that address the impact of the existing fields on the geodesic structure of timelike particles in gravity theories   ~\cite{Shaymatov22c,Dadhich22a,Shaymatov21d,Shaymatov21pdu,aa21}. With this motivation, in this paper, we consider a RN-like BH spacetime with its line element, as described by Eq.~(\ref{m1}) and we further investigate its important insights such as the null and timelike geodesic structures as well as the ISCO and possible particle trajectories. For this geometry, we also consider the light deflection angle using the weak field limit approach and determine the Lorentz-breaking effect to exhibit striking differences of its background geometry near horizon regions. 

The paper is structured as follows: Section \ref{sec2} provides an introduction to the RN-like spacetime background. Section \ref{sec3} examines the geodesics within RN-like spacetime, discussing the analytical solutions for both null and timelike geodesics. In Section \ref{sec4}, we investigate the motion of charged particles. Section \ref{sec5} aims to elucidate the process of calculating the deflection angle under the weak field approximation for the RN-like BH. Finally, our conclusions are presented in Section \ref{sec6}.

\section{Introduction to spacetime background} \label{sec2}

The line element that describes an electrically charged BHs  in gravity with
a background KR field in the standard Schwarzschild-like
coordinates is \cite{RN1}
\begin{equation} 
ds^{2}=-F\left( r\right) dt^{2}+\frac{dr^{2}}{F\left( r\right) }+r^{2}\left(
d\theta ^{2}+\sin ^{2}\theta d\phi ^{2}\right) ,\label{m1}
\end{equation}
where 
\begin{equation} \label{is0}
F\left( r\right) =\frac{1}{1-b}-\frac{2M}{r}+\frac{Q^{2}}{\left( 1-b\right)
^{2}r^{2}},
\end{equation}%
in which, $M$ is the \ BH mass and $b$ is the dimensionless parameter.
Lorentz-violating effects are characterized by the parameter $b$ in this
spacetime, whose value is constrained to be very small from classical
gravitational experiments within the solar system  such as the perihelion
precession of Mercury, deflection of light, and Shapiro time delay  \cite{st1}.  Based on the results of these experiments, they were able to constrain the Lorentz-violating parameter $b$ (Table \ref{tab:my_label}. 
\begin{table}[]
    \centering
  \begin{tabular}{|c|c|} \hline
        Solar Tests & Constraints \\ \hline
Mercury precession & $-3.7\times 10^{-12}\leq b\leq 1.9\times 10^{-11}$ \\ 
Light deflection & $-1.1\times 10^{-10}\leq b\leq 5.4\times 10^{-10}$ \\ 
Shapiro time-delay & $-6.1\times 10^{-13}\leq b\leq 2.8\times 10^{-14}$ \\ \hline
    \end{tabular}
    \caption{ Constraints on the Lorentz-violating parameter $b$ from Solar System tests \cite{st1}.}
    \label{tab:my_label}
    \label{tab:my_label}
\end{table}
   
\begin{figure}
    \centering
    \includegraphics{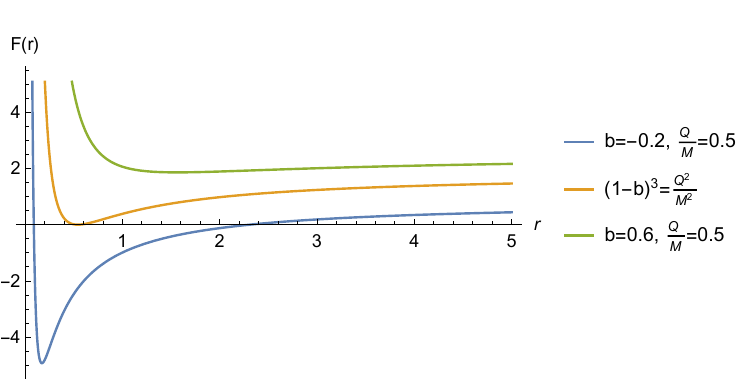}
    \caption{The plots of $F(r)$ for different values of the Lorentz violation parameter $b$}
    \label{fig1}
\end{figure}

 When $b=0$ we recover RN BH
metric. The RN-like spacetime metric (\ref{m1}) has two horizons 
\begin{equation}
r_{\pm }=\left( 1-b\right) \left( M\pm \sqrt{M^{2}-\frac{Q^{2}}{\left(
1-b\right) ^{3}}}\right) .
\end{equation}
For the existence of the horizons the condition $Q^{2}/M^{2}\leq \left(
1-b\right) ^{3}$ is satisfied. The equality represents the case of extreme
BHs. It is important to note that all numerical values used in this study meet the condition above. The  $r_{+}$ horizon corresponds to Schwarzschild's $r = 2M$ event horizon, while the $r_{-}$ horizon is known as the Cauchy horizon. To clarify the impact of $b$ we make a plot of the metric
function as shown in Fig. \ref{fig1}. The Figure depicts that the metric (\ref{m1}) supports three different BH solutions: BH, non-BH, and extreme BH.  Moreover, a brief study of the associated scalars gives a brief insight into the RN-like BH's properties. Hence, \begin{equation}
 R=\frac{2b}{\left( b-1\right) r^{2}},
\end{equation}%
\begin{equation}
R_{\mu \nu }R^{\mu \nu }=\frac{4Q^{4}+4\left( b-1\right) bQ^{2}r^{2}+2\left(
b-1\right) ^{2}b^{2}r^{4}}{\left( b-1\right) ^{4}r^{8}},
\end{equation}%
 \begin{align}\label{RRR}
R_{\mu \nu \alpha \beta }R^{\mu \nu \alpha \beta }& =\frac{4 (b-1)^2 r^2 \left(b^2 r^2+12 (b-1)^2 M^2+4 (b-1) b M r\right)-8 (b-1) Q^2 r (12 (b-1) M+b r)}{(b-1)^4 r^8}\nonumber\\
&+\frac{56 Q^4}{(b-1)^4 r^8}
\end{align}
This demonstrates that the associated scalars are dependent on the Lorentz violating parameter $b$. However, the Ricci scalar is linearly dependent, whereas the Ricci squared and Kretchmann scalars are nonlinear.
\section{The geodesics in RN-like spacetime}\label{sec3}

To study the geodesics in the RN-like BH, we employ the Euler-Lagrange equation given
by \begin{equation}
\frac{d}{ds}\left( \frac{\partial \mathit{L}}{\partial \overset{\cdot }{%
x^{\mu }}}\right) =\frac{\partial \mathit{L}}{\partial x^{\mu }},
\end{equation}%
where $s$ is the affine parameter of the light trajectory, a dot denotes a
differentiation with respect to $s$ and $\overset{\cdot }{x^{\mu }}$
represents the four-velocity of the light ray. The Lagrangian for the
spacetime described in (\ref{m1}) is given by
\begin{equation}
\mathit{L}=\frac{1}{2}\left[ -F\left( r\right) \dot{t}^{2}+\frac{\dot{r}^{2}%
}{F\left( r\right) }+r^{2}\left( \dot{\theta}^{2}+\sin ^{2}\theta \dot{\phi}%
^{2}\right) \right] .
\end{equation}
Since the metric coefficients do not explicitly depend on time $ t$ and azimuthal angle $\phi$, there are two constants of motion which can be labeled as $E$ and $%
\ell $, therefore%
\begin{equation}
E=-F\left( r\right) \dot{t},  \label{en1}
\end{equation}%
and 
\begin{equation}
\ell =r^{2}\sin ^{2}\theta \dot{\phi},  \label{ang1}
\end{equation} where $E$ and $\ell$ are, respectively, the energy and the angular momentum
and of the particle. 
Furthermore, we require $\theta
=\pi /2$ and $\overset{\cdot }{\theta }=0,$ indicating that the particle always moves in the equatorial plane.  With two constants of motion
given in Eqs. (\ref{en1}) and 
 (\ref{ang1}) the geodesics equation becomes 
\begin{equation}
\left( \frac{dr}{d\phi }\right) ^{2}=\frac{r^{4}}{\ell ^{2}}\left(
E^{2}-\left( \frac{1}{1-b}-\frac{2M}{r}+\frac{Q^{2}}{\left( 1-b\right)
^{2}r^{2}}\right) \right) \left( \epsilon +\frac{\ell ^{2}}{r^{2}}\right) .
\end{equation}%
We can obtain the $r$ equations as a function of $s$ and $t$ as follows%
\begin{equation}
\left( \frac{dr}{ds}\right) ^{2}+2V_{eff}\left( r\right) =E^{2},
\end{equation}%
\begin{equation}
\left( \frac{dr}{dt}\right) ^{2}=\left( 1-\frac{2}{E^{2}}V_{eff}\left(
r\right) \right) \left( \frac{1}{1-b}-\frac{2M}{r}+\frac{Q^{2}}{\left(
1-b\right) ^{2}r^{2}}\right) ^{2}.
\end{equation}%
where the effective potential $V_{eff}\left( r\right) $ can be defined as 
\begin{equation}
V_{eff}\left( r\right) =\frac{1}{2}\left( \frac{1}{1-b}-\frac{2M}{r}+\frac{%
Q^{2}}{\left( 1-b\right) ^{2}r^{2}}\right) \left( \epsilon +\frac{\ell ^{2}}{%
r^{2}}\right) .
\end{equation}%
To study the orbits, we make the change of variable $u=1/r$ and obtain,%
\begin{equation}
\left( \frac{du}{d\phi }\right) ^{2}=\frac{E^{2}}{\ell ^{2}}-\left( \frac{1}{%
1-b}-2Mu+\frac{Q^{2}}{\left( 1-b\right) ^{2}}u^{2}\right) \left( \frac{%
\epsilon }{\ell ^{2}}+u^{2}\right) =g\left( u\right) .\label{gu1}
\end{equation}%
We choose $\left( \epsilon =0\right) $ for null and $\left( \epsilon
=1\right) $ for timelike geodesic.

\subsection{The null geodesics}

In this part we move now to investigate null geodesics of particle in RN-like BH spacetime described by Eq. (\ref{m1}). As was mentioned above for the 4-momentum $p_\mu p^\mu=\epsilon$ we have  set $\epsilon=1$ for massive particle while $\epsilon=0$ for null geodesics. 

For null geodesics we rewrite the radial equations of motion photons in the following form as 
\begin{eqnarray} \label{Veff3}
\dot{r}^2=E^2+\epsilon F(r)-\frac{\ell^2}{r^2}F(r)
\, ,
\end{eqnarray}
where $\epsilon=1$ can for simplicity be set for timelike geodesics so that one describes energy $E$ and angular momentum $\ell$ per unit mass. To find the radii of circular orbits for given values of $E$ and $L$ one can solve $\dot{r}=\ddot{r}=0$ simultaneously, i.e.,
\begin{eqnarray}
V_{eff}(r,E,\ell)=0, \mbox{~~~} \frac{\partial V_{eff}(r,E,\ell)}{\partial r}=0\, ,
\end{eqnarray}
where $V_{eff}(r,E,\ell)$ can be given by
\begin{eqnarray}
V_{eff}(r,E,\ell)&=& E^2r^2+\epsilon\,r^2F(r)-\ell^2F(r)\, .
\end{eqnarray}

For the null geodesics the radius of $r_{ph}$ can be determined by the angular momentum's minimum value, $\ell=\ell(r)$, which is obtained by solving $V_{eff}'=0$. For null geodesics using $V_{eff}'=0$ is sufficient so that one can find the following condition for the photon sphere radius
\begin{eqnarray}
\frac{1}{1-b}-\frac{3M}{r_{ph}}+\frac{2Q^{2}}{\left( 1-b\right) ^{2}\,r_{ph}^{2}} =0\, ,
\end{eqnarray}
which solves to give the photon orbit $r_{ph}$ implicitly as 
\begin{eqnarray}
r_{ph}=\frac{3 }{2}(1-b) M+\frac{\sqrt{(1-b) \left(9 (1-b)^3 M^2-8 Q^2\right)}}{2 (1-b)}\, .
\end{eqnarray}
It is obvious that the photon sphere retrieves the Schwarzschild case in the limit of $b=0$ and $Q=0$, i.e., $r_{ph}=3M$. 
\begin{figure}
    \centering    \includegraphics[width=10.0cm]{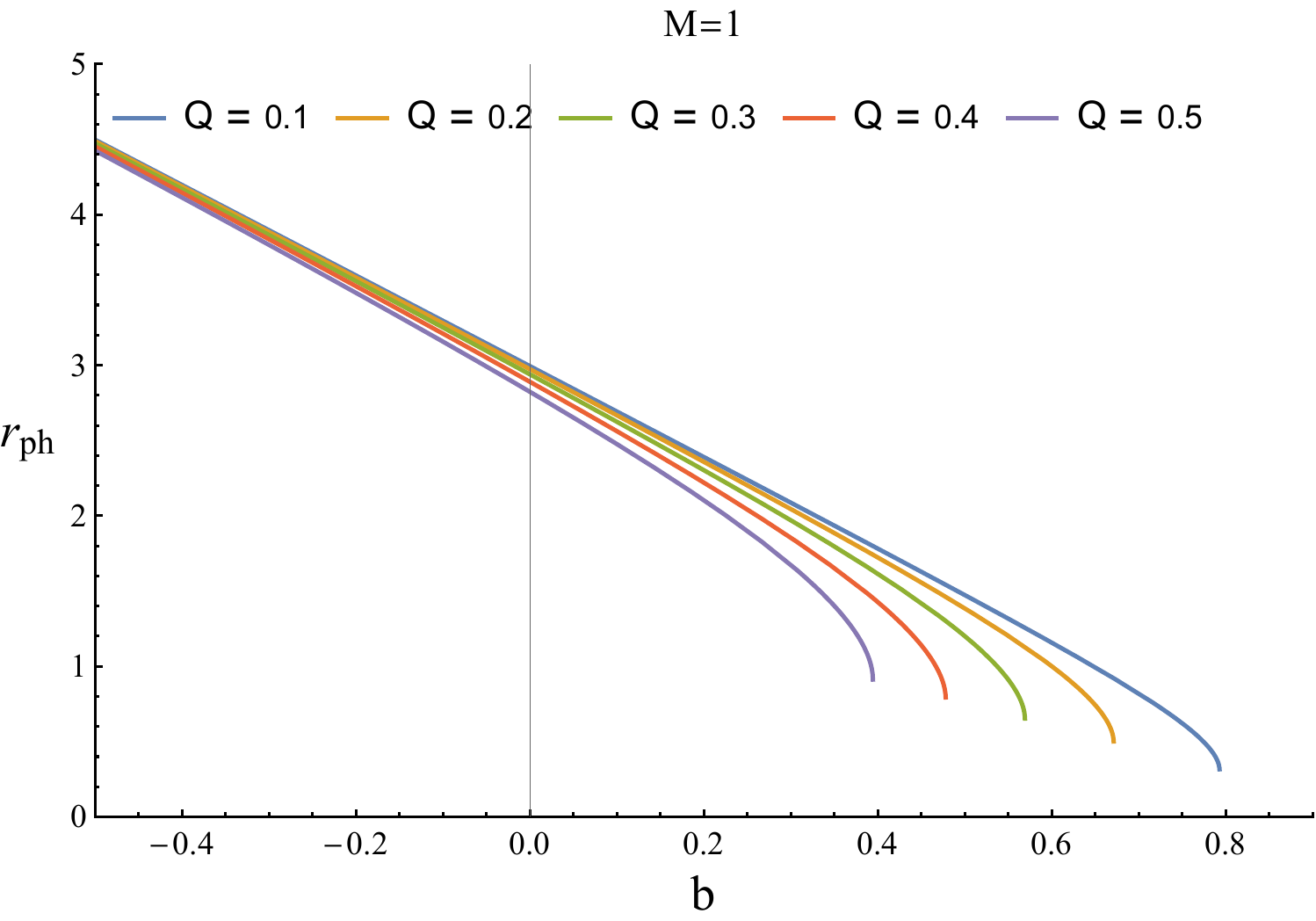}
    \caption{The profile of the photon sphere for various values of BH charge $Q$.}
    \label{fig_ph}
\end{figure}
In Fig.~\ref{fig_ph},  we show the dependence of the photon radius from the parameter $b$ for the various values of black hole charge $Q$. As can be observed from Fig.~\ref{fig_ph}, the photon radius decreases up to its minimum value as the parameter $b$ increases. Similarly, one can observe the same behaviour for black hole charge $Q$, that is the photon radius also decreases gradually. It is to be emphasized that the presence of the parameter $b$ has the physical effect of shifting the photon sphere inwards to the central singularity. The physical effect of parameter $b$ can therefore be interpreted as a repulsive gravitational charge, similarly to what is observed for black hole electric charge $Q$.

\subsection{ Timelike geodesics}

The equations describing the radial timelike geodesics are $(\ell=0$ and $\epsilon=0)$ \begin{equation}
\left( \frac{dr}{ds}\right) ^{2} =E^{2}-\left(\frac{1}{1-b}-\frac{2M}{r}+\frac{Q^{2}}{\left( 1-b\right)
^{2}r^{2}}\right),
\end{equation}%
\begin{equation}
     \frac{dt}{d\tau} =\frac{E}{\frac{1}{1-b}-\frac{2M}{r}+\frac{Q^{2}}{\left( 1-b\right)
^{2}r^{2}}}.
\end{equation}
Because $\frac{1}{1-b}-\frac{2M}{r}+\frac{Q^{2}}{\left( 1-b\right)
^{2}r^{2}}$ is positive in the interval $0\leq r<r_{-}$, then the trajectory will have a turning point inside the Cauchy horizon. Horizon singularities differ between RN-like and Schwarzschild geometries. The former provides more insight into the nature of trajectories as it is timelike, while the latter is spacelike. Particles can move from one spacetime region to another by crossing the $r_{+}$ horizon and skirting the singularity. In contrast, in Schwarzschild geometry, after crossing the event horizon at $r = 2M$, the particle must fall towards the singularity. 

\begin{figure}
    \centering    \includegraphics{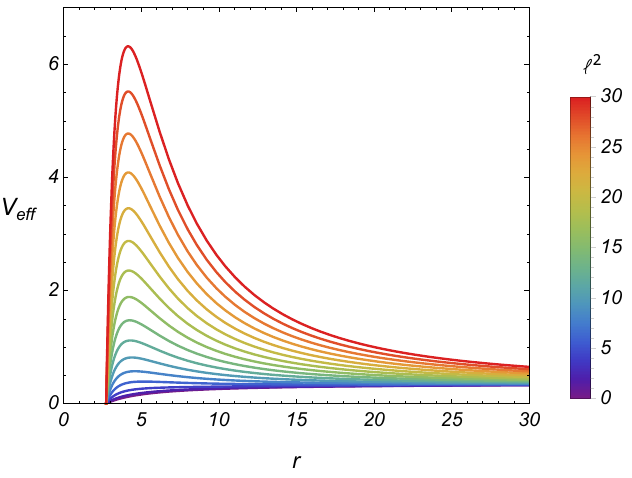}
    \caption{The profile of the effective potential  (\ref{effpot}) curves of timelike particles for different values of $\ell^2$. Here, $M=1, Q=0.4$ and $b=-0.4$}
    \label{figeffpot1}
\end{figure}

The effective potential  for timelike particles $(\epsilon=1)$ is given by \begin{equation}
  V_{eff}\left( r\right) =\frac{1}{2}\left( \frac{1}{1-b}-\frac{2M}{r}+\frac{%
Q^{2}}{\left( 1-b\right) ^{2}r^{2}}\right) \left( 1 +\frac{\ell ^{2}}{%
r^{2}}\right) . \label{effpot} 
\end{equation} The effective potential (\ref{effpot}) depends on the BH parameters and the angular momentum $\ell^2$. The plot of the effective potential for timelike particles for different values of $\ell^2$ is shown in Figure \ref{figeffpot1}. The Figure demonstrates that the timelike orbits in RN-like BH are all unstable circular orbits, as the effective potential curves corresponding to different values of $\ell^2$ all exhibit only one maximum point. \\The condition for the occurrence of circular orbits are  $g(u)=0$ and $g'(u)=0$. From these conditions, it follows that the energy and the angular momentum of a circular  orbit of radius $r_c=1/u_c$ is given by 
\begin{figure*}
\centering
\includegraphics[width=1.0\textwidth]{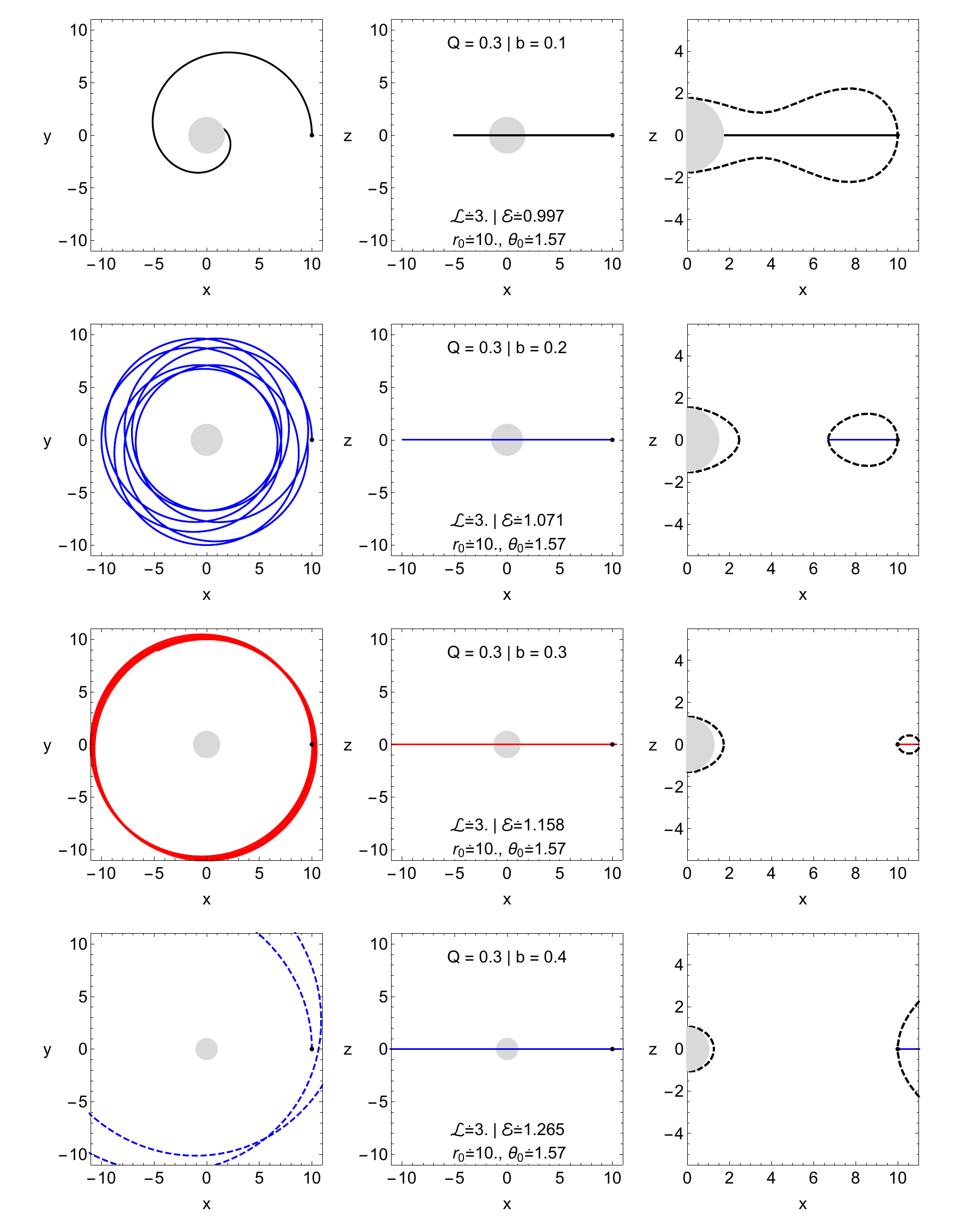}%

\caption{\label{fig:traj1} Plot shows trajectories of the timelike neutral particles for the polar plane (i.e. $z = 0$, the first column) and the equatorial plane (i.e. $y = 0$, the second column) and the boundaries of neutral particle motion for the equatorial plane (i.e. $y = 0$, the third column) around BHs in KR gravity field for various possible cases.} 
\end{figure*}
\begin{equation}
E^{2}=\frac{\left( \frac{1}{1-b}-2Mu_{c}+\frac{Q^{2}}{\left( 1-b\right) ^{2}}%
u_{c}^{2}\right) ^{2}}{D}, \label{en3}
\end{equation}
\begin{equation}
\ell ^{2}=\frac{M-\frac{Q^{2}}{\left( 1-b\right) ^{2}}u_{c}}{Du_{c}},  \label{ang3}
\end{equation}
where $D=\frac{1}{1-b}-3Mu_{c}+\frac{2Q^{2}}{\left( 1-b\right) ^{2}}%
u_{c}^{2}>0$. Accordingly, with $ d^2V_{eff}/dr^2= 0$ and Equations (\ref{en3}-\ref{ang3}), we can solve for the radius of ISCO \cite{RN1}. Beside this circular orbit, Eq. (\ref{gu1}) with $\epsilon=1$ provides an orbit of the second kind determined by  
\begin{equation}
\phi =\pm \int \frac{du}{\sqrt{g(u)}}, \label{phi2}
\end{equation}
where 
\begin{equation}
g\left( u\right) =\left( u-u_{c}\right) ^{2}\left[\frac{Q^{2}}{\left( 1-b\right)
^{2}}
(u^{2}-u_{c}^{2})+2\left( M-\frac{Q^{2}}{\left( 1-b\right)
^{2}} u_{c}\right) u+Mu_{c}-\frac{M}{\ell^2 u_c}\right] .
\end{equation}%
Substitute $ \xi =\frac{1}{u-u_{c}}$ we obtain the solution of (\ref{phi2}) as 
\begin{equation}
\phi =\pm \frac{1}{A}\ln \left( 2A\xi +B+2A\sqrt{A\left( \frac{Q^{2}}{\left( 1-b\right)
^{2}} +B\xi +A\xi ^{2}\right) }\right) ,
\end{equation}%
where 
\begin{equation}
A=3M u_c-4\frac{Q^{2}}{\left( 1-b\right)
^{2}}u^2_c -\frac{M}{\ell^2 u_c} 
\end{equation} and
\begin{equation}
B=2\left(M-2\frac{Q^{2}}{\left( 1-b\right)
^{2}} u_c\right).
\end{equation}

To gain a deeper understanding qualitatively how timelike geodesics can behave around the RN-like BH in the presence of KR field we further demonstrate a timelike particle trajectory as a consequence of the combined effects of parameters $b$ and $Q$. It should be noted that we shall for simplicity restrict timelike particle trajectory to the equatorial plane of the RN-like BH. In doing so, we show the timelike particle trajectory in Fig. \ref{fig:traj1}. It is clearly seen that the timelike particle orbits remain the captured orbits in the case of small values of parameter $b$, while these orbits gradually become the bounded as it increases. However, the orbits of the timelike particles can turn into the escaping orbits for overestimated values of parameter $b$ in the case in which the energy and angular momentum of the timelike particle are fixed. This happens because the repulsive gravitational impact on the particle trajectory grows under the combined effects of parameters $b$ and $Q$, thus resulting in having the escaping orbits.  

\subsection{Astrophysical applications}\label{apllications}

We now turn to consider astrophysical applications. It has value to note that although the recent modern experiments and observations that pertain to gravitational waves \cite{Abbott16a,Abbott16b} and the first image of the galaxy M87 as supermassive BH candidate~\cite{Akiyama19L1,Akiyama19L6} have proven the existence of astrophysical BHs in the universe, yet they have remained BH candidates as that of no exact departures, leaving an open window for the precise measurements of BH parameters such as mass, angular momentum, etc. It does therefore make it challenging to identify the unique signatures of possible BH candidates as viable sources from theoretical and observational perspectives. With this motivation in mind, we consider the repulsive impacts of Lorentz-violating parameter $b$ that can, as discussed above, affect the observable quantities thus having the behaviour of geodesic structure of RN-like BH in gravity with a background KR field, which is similar in contrast with one for a rotating Kerr BH spacetime. Hence, the effect of Lorentz-violating parameter would be able to mimic the effects of BH spin parameter and would play an important role for testing observable properties, i.e., the innermost stable circular orbit (ISCO). For being more illustrative, we consider stable circular orbits and find the relation between BH spin parameter $a$ of the Kerr spacetime and Lorentz-violating parameter $b$ of RN-like BH spacetime in gravity with a background KR field. For that we compare the ISCO for both spacetime geometries. 
According to Bardeen \textit{et al.}~\cite{Bardeen72} it is more illuminating to write the ISCO expression of massive test particles moving around Kerr BH, and it is given by 
\begin{eqnarray}
r_{\rm ISCO}= 3 + Z_2 \pm \sqrt{(3- Z_1)(3+ Z_1 +2 Z_2 )} \ ,
\end{eqnarray}
with
\begin{eqnarray} 
Z_1 = 
1+\left( \sqrt[3]{1+a}+ \sqrt[3]{1-a} \right) 
\sqrt[3]{1-a^2} \mbox{\qquad and \qquad}
Z_2 = \sqrt{3 a^2 + Z_1^2}\, .
\end{eqnarray}
\begin{figure*}
    \centering    
    \includegraphics[width=8.0cm]{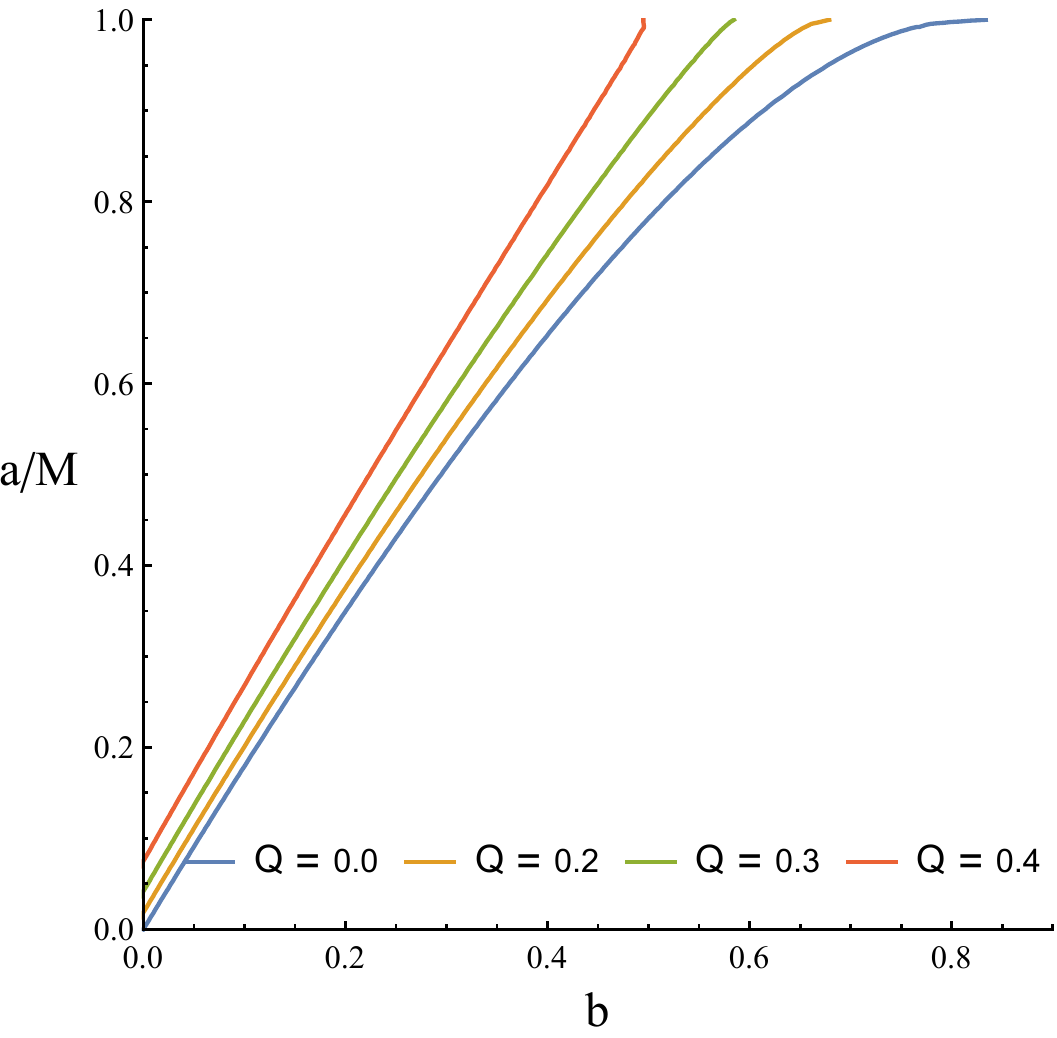}
    \includegraphics[width=8.0cm]{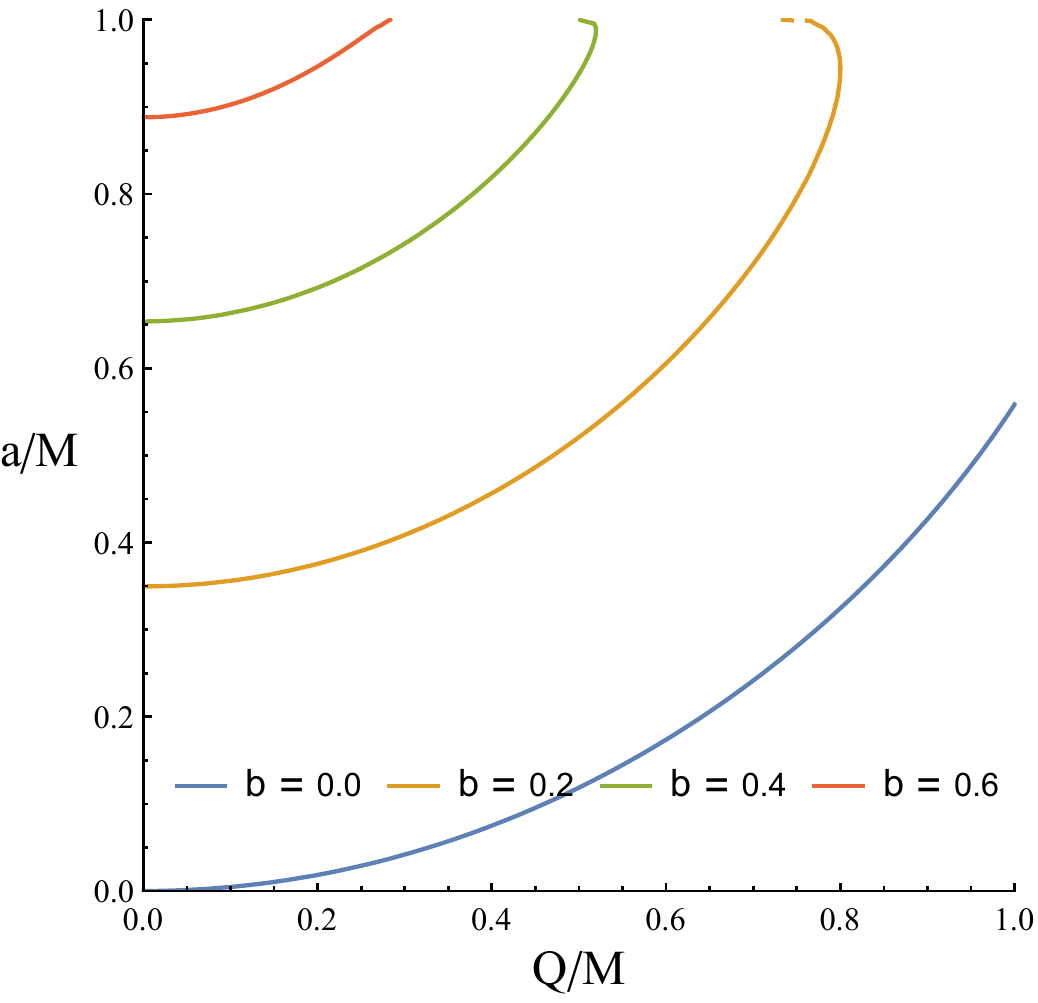}
    \caption{The profile of the values of spin parameter $a/M$ as a function of Lorentz-violating parameter $b$ and BH charge $Q/M$ for which the ISCO radius for the Kerr BH is identical as the ISCO for RN-like BH in gravity with a background KR field.}
    \label{fig_mimic}
\end{figure*}

To this end, in Fig.~\ref{fig_mimic} we show the values of spin parameter $a$ as a function of $b$ and $Q$ for which the ISCO radius for RN-like BH is similar to the one for the Kerr BH spacetime. It is clearly seen from Fig.~\ref{fig_mimic} that the corresponding values of $b$ and $Q$ refer to the exact values of spin parameter $a$ of Kerr BH. The effect of Lorentz-violating parameter $b$ can give similar effect with a larger values of Kerr BH spin parameter $a$, i.e., the geodesics of particles around a RN-like BH could be the same as the geodesics around a rotating Kerr BH. Interestingly, it can be observed that the combined effect of $b$ and $Q$ can expedite the mimicking process, as seen in Fig.~\ref{fig_mimic}. That is, the combined effects of Lorentz-violating parameter and BH charge can mimic the BH spin parameter up to its maximum values, i.e., $a/M \approx 1$. The main concern to be noted is that a distant observer would not able to differentiate between two various BH geometries as stated by the analysis of the emitted electromagnetic radiations from BH accretion disk. With this in regard, the results suggest that the presence of Lorentz-violating parameter alone can affect a measurement of BH spin parameter up to 100\%. One can then infer from the above qualitative findings that the current 
precision of measurements may not rule out the Lorentz-violating parameter due to the fact that rapidly rotating astrophysical BH candidates can be considered as RN-like BH in gravity with a background KR field~\cite{Walton13,Patrick11b-Seyfert,Patrick11a-Seyfert,Bambi17-BHs}.

\subsection{Extracted energy by collisions of timelike particles}

\begin{figure*}
    \centering    
    \includegraphics[width=8.0cm]{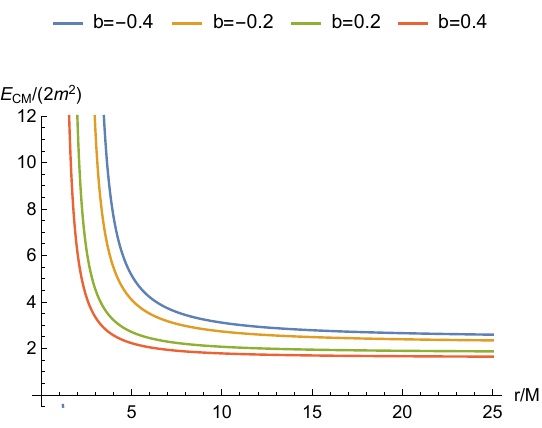}
    \includegraphics[width=8.0cm]{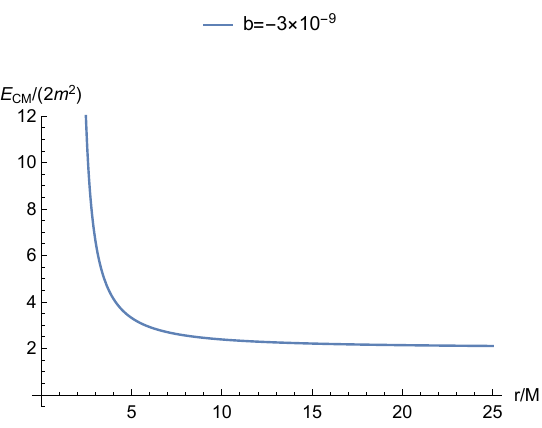}
    \caption{The variation of the CM energy $E_{CM}$ of the colliding neutral particles for the fixed $Q$ and various values of $b$ (left) and constraint value (right).}
    \label{exEn}
\end{figure*}

In this part, we investigate the acceleration of charged particles colliding near the BH horizon under an electrically charged BHs in gravity with a background KR gravity. We analyze two particles with the same masses, $m_{1}=m_{2}=m$, at a distance far from BH, and we can write equations of motion like\begin{equation}
    P^\mu=mu^\mu,\hspace{1cm}u^t=\frac{E}{F(r)}, \label{ex1}
\end{equation}\begin{equation}
   u^\phi=\frac{\ell}{r^2},\hspace{1cm}u^r=\sqrt{E^2- F(r)\left( 1 +\frac{\ell ^{2}}{%
r^{2}}\right)}. \label{ex2}
\end{equation}
The extracted energy in the center of mass
frame for this collision is defined as \cite{ene1} \begin{equation}
    \frac{E_{CM}}{2m^2}=1-g_{\mu\nu}u^{\mu}_1u^{\nu}_2. \label{ex3} 
\end{equation} When substituting  Eqs. (\ref{ex1}) and (\ref{ex2}) into Eq. (\ref{ex3}) we obtain  
\begin{equation}
    \frac{E_{CM}}{2m^2}=1+\frac{E_1E_2}{F(r)}-\frac{\ell_1 \ell_2}{r^2}- \frac{1}{F^2(r)}\sqrt{E_1^2- F(r)\left( 1 +\frac{\ell_1 ^{2}}{%
r^{2}}\right)}\sqrt{E_2^2- F(r)\left( 1 +\frac{\ell_2 ^{2}}{%
r^{2}}\right)}
. \label{ex4} 
\end{equation}
In Figure \ref{exEn}, the CM energy of these two colliding particles is plotted against the radius of circular orbit according to Eq. (\ref{ex4}). The Lorentz-violating effect is clearly seen in Figure \ref{exEn} where the CM energy of colliding test particles decreases as the Lorentz-violating parameter increases. 

\section{Motion of a charged particle} \label{sec4}

\begin{figure}
    \centering
{{\includegraphics[width=7.5cm]{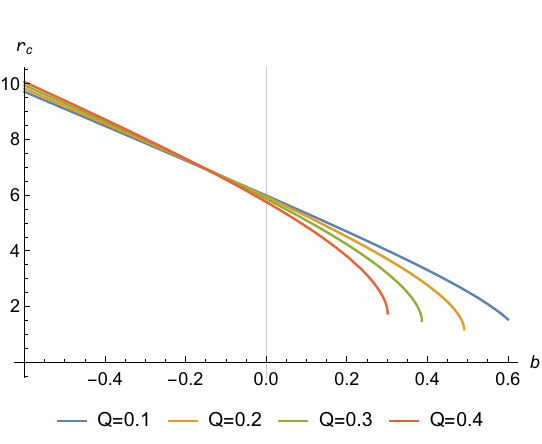} }}\qquad
    {{\includegraphics[width=7.5cm]{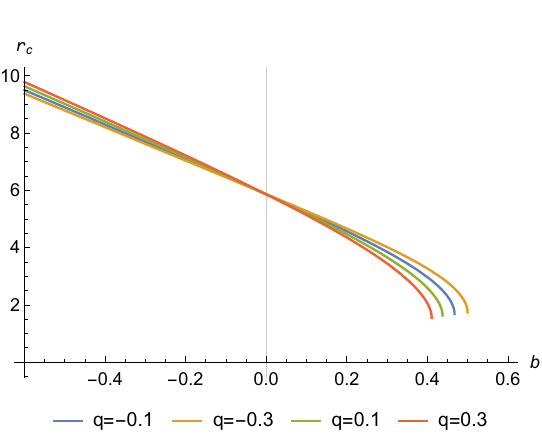}}}
    \caption{ Plot of circular radius (\ref{rad5}) with varying values of $Q$ and $q$.}
    \label{figangular}
\end{figure}

In this section we will study the circular motion of a test particle which
has a charge per unit mass $q$, in the vicinity of a BH in gravity with a background
KR field. Recall that metric (\ref{m1}) has an electric scalar potential of
the form $\frac{Q}{\left( 1-b\right) r}$ \cite{RN1} hence, the Lagrangian 
\begin{equation}
2\mathit{L}=-F\left( r\right) \dot{t}^{2}+\frac{1}{F\left( r\right) }\dot{r}%
^{2}+r^{2}\left( \dot{\theta}^{2}+\sin ^{2}\theta \dot{\phi}^{2}\right) +%
\frac{qQ}{\left( 1-b\right) r}\dot{t}.
\end{equation}%
The trajectory equation for the charged particle with mass $m$ is,%
\begin{equation*}
\left( \frac{du}{d\phi }\right) ^{2}=-\frac{Q^{2}}{\left( 1-b\right) ^{2}}%
u^{4}+2Mu^{3}-\left( \frac{\left( 1-b\right) ^{2}\mathit{\ell }%
^{2}+Q^{2}\left( 1-q^{2}\right) }{\left( 1-b\right) ^{2}\mathit{\ell }^{2}}%
\right) u^{2}
\end{equation*}%
\begin{equation}
+\frac{1}{\mathit{\ell }^{2}}\left( M-\frac{qQ\mathit{E}}{\left( 1-b\right) m%
}\right) u+\frac{\mathit{E}^{2}-1}{\mathit{\ell }^{2}}=g\left( u\right) ,
\end{equation}%
where $\mathit{E}=E/m$ and $\mathit{\ell }=\ell /m$. The conditions for the
occurrence of circular orbits are $g\left( u\right) =0$ and 
\begin{equation*}
g^{\prime }\left( u\right) =-4\frac{Q^{2}}{\left( 1-b\right) ^{2}}%
u^{3}+6Mu^{2}-2\left( \frac{\left( 1-b\right) ^{2}\mathit{\ell }%
^{2}+Q^{2}\left( 1-q^{2}\right) }{\left( 1-b\right) ^{2}\mathit{\ell }^{2}}%
\right) u
\end{equation*}%
\begin{equation}
+\frac{1}{\mathit{\ell }^{2}}\left( M-\frac{qQ\mathit{E}}{\left( 1-b\right) m%
}\right) =0
\end{equation}%
The expressions for energy $\mathit{E}$ and angular momentum $\mathit{\ell }$
of a circular orbit of radius $r_{c}=1/u_{c}$ can be obtained from the above
two conditions hence, 
\begin{equation}
E^{2}=\frac{\left( \frac{1}{1-b}-2Mu_{c}+\frac{Q^{2}}{\left( 1-b\right) ^{2}}%
u_{c}^{2}\right) ^{2}}{D}+\frac{\frac{qQ\mathit{E}}{\left( 1-b\right) m}\left[ \frac{1}{1-b}-4Mu_{c}+3%
\frac{Q^{2}}{\left( 1-b\right) ^{2}}u_{c}^{2}\right] u_{c}}{D} \label{en2}
\end{equation}%
\begin{equation*}
+\frac{q^{2}\frac{Q^{2}}{\left( 1-b\right) ^{2}}\left[ 1-\frac{Q^{2}}{\left(
1-b\right) ^{2}}u_{c}\right] u_{c}^{3}}{D}
\end{equation*}%
\begin{equation}
\ell ^{2}=\frac{M-\frac{Q^{2}}{\left( 1-b\right) ^{2}}u_{c}-\frac{qQ}{\left(
1-b\right) }\left( \frac{\mathit{E}}{m}-\frac{qQ}{\left( 1-b\right) }%
u_{c}\right) }{Du_{c}}.  \label{ang2}
\end{equation}%
Recall that, the minimum radius for a stable circular orbit occurs at the
turning point of the function $g(u)$, hence%
\begin{equation}
g^{\prime \prime }\left( u\right) =-6\frac{Q^{2}}{\left( 1-b\right) ^{2}}%
u^{2}+6Mu-\left( \frac{\left( 1-b\right) ^{2}\mathit{\ell }^{2}+Q^{2}\left(
1-q^{2}\right) }{\left( 1-b\right) ^{2}\mathit{\ell }^{2}}\right) =0.
\label{g1}
\end{equation}%
Using Eq. (\ref{ang2}) to eliminate $\ell ^{2}$ then Eq. (\ref{g1}) becomes%
\begin{equation*}
4\frac{Q^{4}}{\left( 1-b\right) ^{4}}u_{c}^{3}-9M\frac{Q^{2}}{\left(
1-b\right) ^{2}}u_{c}^{2}-M-
\end{equation*}%
\begin{equation}
q\frac{Q}{\left( 1-b\right) }\left[ q\frac{Q}{\left( 1-b\right) }\left( 4%
\frac{Q^{2}}{\left( 1-b\right) ^{2}}u_{c}^{3}-3Mu_{c}^{2}\right) -\frac{E}{m}%
\left( \frac{6Q^{2}}{\left( 1-b\right) ^{2}}u_{c}^{2}-6Mu_{c}+1\right) %
\right] =0
\end{equation}%
or in terms of $r_{c}$%
\begin{equation}
r_{c}^{3}-6\left( 1-b\right) Mr_{c}^{2}+9\frac{Q^{2}}{\left( 1-b\right) ^{2}}%
r_{c}-\frac{4Q^{4}}{\left( 1-b\right) ^{4}M}-
\end{equation}%
\begin{equation*}
\frac{qE}{m}\frac{Q}{\left( 1-b\right) }\left( \frac{r_{c}^{3}}{M}%
-6r_{c}^{2}+\frac{6Q^{2}}{\left( 1-b\right) ^{2}M}r_{c}\right) -q^{2}\frac{%
Q^{2}}{\left( 1-b\right) ^{2}}\left( 3r_{c}-\frac{4Q^{2}}{\left( 1-b\right)
^{2}M}\right) =0.
\end{equation*}%
Ignoring terms of order $\frac{Q^{4}}{\left( 1-b\right) ^{4}}$we obtain
\begin{equation}
r_{c}=\frac{6\left( 1-b\right) M-\frac{6qE}{m}\frac{Q}{\left( 1-b\right) }}{%
2\left( 1-\frac{qE}{m}\frac{Q}{\left( 1-b\right) }\right) }\pm   \label{rad5}
\end{equation}%
\begin{equation*}
\frac{\sqrt{\left( 6\left( 1-b\right) M-\frac{6qE}{m}\frac{Q}{\left(
1-b\right) }\right) ^{2}-4\left( 1-\frac{qE}{m}\frac{Q}{\left( 1-b\right) }%
\right) \left( 9\frac{Q^{2}}{\left( 1-b\right) ^{2}}-9\frac{q^{2}Q^{2}}{%
\left( 1-b\right) ^{2}}-\frac{6qE}{m}\frac{Q^{3}}{\left( 1-b\right) ^{3}M}%
\right) }}{2\left( 1-\frac{qE}{m}\frac{Q}{\left( 1-b\right) }\right) }.
\end{equation*}
Note that, for $Q=0,$ then (\ref{rad5}) reduces to 
\begin{equation}
r_{c}=6\left( 1-b\right) M.
\end{equation}%
which is the  ISCO radius of the Schwarzschild-like BH.  The circular
orbits for charged particle (\ref{rad5})  is shown in Figure \ref{figangular}.
\begin{figure*}
\centering
\includegraphics[width=1.0\textwidth]{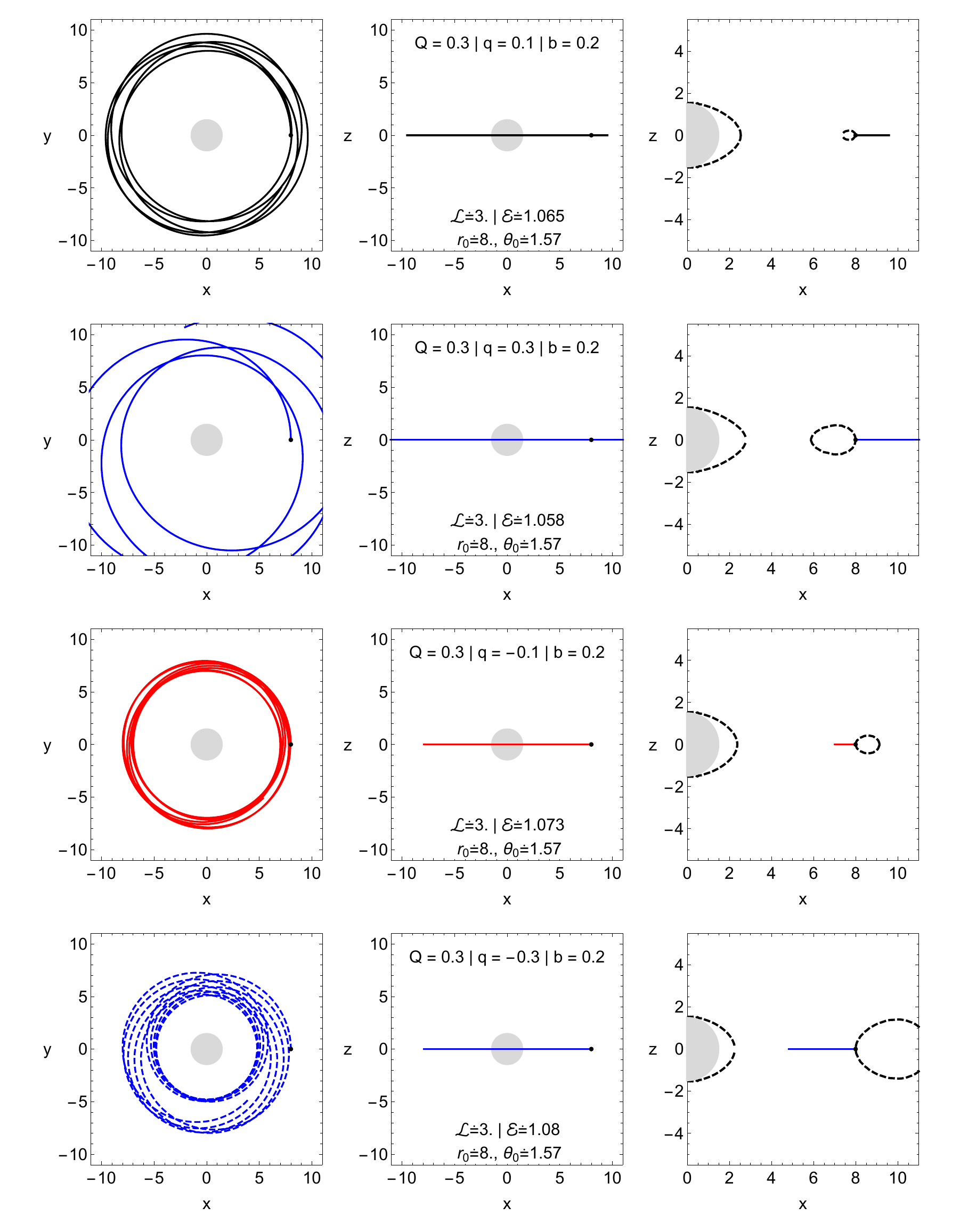}%
\caption{\label{fig:traj2} Plot shows trajectories of the timelike charged particles for the polar plane (i.e. $z = 0$, the first column) and the equatorial plane (i.e. $y = 0$, the second column) and the boundaries of charged particle motion for the equatorial plane (i.e. $y = 0$, the third column) around BHs in 
KR gravity field for various possible cases.}
\end{figure*}
It is evident from Figure that the circular radius of an RN-like BH shrinks with increasing Lorentz violating parameter $b$ and charge $Q$. Figure \ref{figangular} illustrates that the ISCO radius is more responsive to changes in the Lorentz-violating parameter $b$ than to changes in the electric charge $Q$. Similarly to what is observed in the trajectory of timelike neutral particle, we further demonstrate how the inclusion of a charged particle can affect on the behaviour of particle trajectory while keeping fixed Lorentz-violating parameter $b$ and charge $Q$; see Fig.~\ref{fig:traj2}. It is clearly seen from Fig.~\ref{fig:traj2} that the charged particle trajectory can be influenced drastically by the combined effects of Coulomb and gravitational forces, thus leading to the change in the behaviour of orbits around the RN-like BH. It turns out that the Coulomb repulsive force dominates over gravitational force that arises from Lorentz-violating parameter $b$, thus making the stable orbits become the escaping orbits in the case of a positive electric charge $q>0$.  However, the opposite is the case for a charged particle with negative $q<0$, i,e., the timelike charged particle orbits remain the captured ones under the combined effect of attractive Coulomb and gravitational forces, as seen in Fig.~\ref{fig:traj2}.

\section{Deflection angle of RN-like BH\lowercase{s} WITH KR FIELDS} \label{sec5}

This section is dedicated to elucidating the process of computing the deflection angle \cite{isOvgun:2018tua} within the framework of gravity involving a background KR field. The calculation is conducted under the weak field approximation for an electrically charged and massive BH, resembling the spherically static configuration. This analysis occurs in a non-plasma medium and utilizes the Gauss-Bonnet theorem (GBth) \cite{isGibbons:2008rj,isOvgun:2018ran,isMandal:2023eae,isBarman:2024hwd}. The GBth assumes a pivotal role in bridging the intrinsic geometry of the metric with the underlying topology within the region $\mho_ {\cal R}$, delineated by its boundary $\partial \mho_ {\cal R}$. This profound connection is succinctly articulated through the following equation \cite{isGibbons:2008rj}:

\begin{equation}\label{is1}
\iint_{\mho_{\cal R}} \mathcal{K} d S_{2D}+\oint_{\partial \mho_{\cal R}} \mathfrak{h} d t+\sum_{z} \alpha_{z}=2 \pi  {\Upsilon}\left(\mho_{\cal {R}}\right),
 \end{equation}
where $\mho_{\cal R}\subset S_{2D}$ is a regular subset of a simple two-dimensional surface $S_{rf}$, characterized by a closed, regular, and positively oriented boundary $\partial \mho_{\cal R}$. In this context, $\mathfrak{h}$ represents the geodesic curvature of $\partial \mho_{\cal R}$, defined as $\mathfrak{h}=\bar{g}\left(\nabla_{\dot{\gamma}} \dot{\gamma}, \ddot{\gamma}\right)$, where $\bar{g}(\dot{\gamma}, \dot{\gamma})=1$, and $\ddot{\gamma}$ denotes the unit acceleration vector. At the $z^{\mbox{th}}$ vertex, $\alpha_{z}$ denotes the exterior angle. Moreover, in Eq. \eqref{is1}, $\Upsilon \left(\mho_{\cal {R}}\right)$ represents the Euler characteristic number \cite{isWang:2022yvi}. Besides, $\mathcal{K}$ represents the Gaussian optical curvature \cite{isUmetsu:2010us}. To determine $\mathcal{K}$, one must consider the null geodesics deflected by the BH \cite{isJusufi:2018kmk}. It is well-established that light follows null geodesics (i.e., $ds^2=0$). These null geodesics are carefully chosen to define the optical metric, which characterizes the Riemannian geometry observed by light. By imposing the null condition together with $\theta
=\pi /2$ (the equatorial plane in the optical metric provides a potential surface of revolution), we reveal the following optical metric in the new coordinate system:
\begin{equation}\label{is4}
dt^2=\tilde{g}_{ij} d\mathrm{x}^i d\mathrm{x}^j= dr_{*}^2+\mathcal{F}^2(r_{*})d\phi^2,
\end{equation}
where 
\begin{equation} \label{is4n}
\mathcal{F}(r_{*}(r))=\frac{r}{\sqrt{F(r)}},
\end{equation}
and $r_{*}$ is named as the tortoise coordinate \cite{isChandrasekhar:1985kt,isPourhassan:2022irk,isSakalli:2016ojo} and it is given by
\begin{equation}\label{is5}
r_{*}=\int{\frac{dr}{F(r)}}=\frac{r}{\alpha}+\frac{M \ln (r^2F(r))}{\alpha^2}-\frac{\left(\xi+M^2\right) \ln \left(\frac{\sqrt{\xi}+\beta}{\sqrt{\xi}-\beta}\right)}{2 \alpha^2 \sqrt{\xi}},
\end{equation}
in which
\begin{eqnarray}\label{is5n}
\alpha=\frac{1}{1-b},\\
\xi=M^2-\alpha^3 Q^2, \\
\beta=\alpha r-M. 
\end{eqnarray}
The non-vanishing Christoffel symbols \cite{isWald:1984rg} associated with metric are computed as:
\begin{eqnarray} \label{is7.1}
\Gamma_{\phi \phi}^{r_{*}}&=&-\mathcal{F}(r_{*})\frac{\mathrm{d}\mathcal{F}(r_{\star})}{\mathrm{d}{r_{\star}}},\\
\Gamma_{r_{*}\phi}^{\phi}&=&\frac{1}{\mathcal{F}(r_{*})}\frac{\mathrm{d}\mathcal{F}(r_{\star})}{\mathrm{d}{r_{\star}}}, \label{is7.2}
\end{eqnarray}
Note that the determinant can be found as $\det\tilde{g}_{ij}=\mathcal{F}^{2}(r^{\star})$. Consequently, one can evaluate the Gaussian optical curvature $\mathcal{K}$ \cite{isMandal:2023eae} as follows:
\begin{equation}
\mathcal{K}=-\frac{R_{r_{*}\phi r_{*}\phi}}{\det\tilde{g}_{r\phi}}=-\frac{1}{\mathcal{F}(r_{\star})}\frac{\mathrm{d}^{2}\mathcal{F}(r_{\star})}{\mathrm{d}{r_{\star}}^{2}}.
\end{equation}

The expression for optical curvature $\mathcal{K}$ can alternatively be reformulated using the variable $r$ \cite{isOvgun:2018ran}. Consequently, one finds out the optical curvature as
\begin{eqnarray}
\mathcal{K} & = &-\frac{1}{\mathcal{F}(r^{\star})}\left[\frac{\mathrm{d}r}{\mathrm{d}r^{\star}}\frac{\mathrm{d}}{\mathrm{d}r}\left(\frac{\mathrm{d}r}{\mathrm{d}r^{\star}}\right)\frac{\mathrm{d}F(r)}{\mathrm{d}r}+\left(\frac{\mathrm{d}r}{\mathrm{d}r^{\star}}\right)^{2}\frac{\mathrm{d}^{2}F(r)}{\mathrm{d}r^{2}}\right].\label{is8}
\end{eqnarray}
After substituting Eqs. \eqref{is0}, \eqref{is4n}, and \eqref{is5} into Eq. \eqref{is8}, the optical curvature is obtained as follows:
\begin{equation} \label{is42}
\mathcal{K} = -2\frac{M\alpha}{{r}^{3}} + 3\frac{{M}^{2}}{{r}^{4}} - 6\frac{{Q}^{2}{\alpha}^{2}M}{{r}^{5}} + 3\frac{{Q}^{2}{\alpha}^{3}}{{r}^{4}} + 2\frac{{Q}^{4}{\alpha}^{4}}{{r}^{6}}.
\end{equation}
One can observe that the Gaussian optical curvature depends on various parameters such as mass $M$, $Q$, and $\alpha$. Now, we would like to determine the deflection angle by employing the GBth. If we take the limit $\cal{R} \rightarrow \infty$ in Eq. \eqref{is1}, the jump angle converges to $\pi/2$, ensuring $\theta_{0}+\theta_{S}=\pi$, and the Euler characteristic number becomes unity  \cite{isJaved:2022bdi}. As a consequence, the following equation can be anticipated:

\begin{equation}\label{is43a}
\iint_{\mho_{\cal R}} \mathcal{K} d S+\oint_{\partial \mho_{\cal R}} \mathfrak{h} d t+\alpha_{z}=2 \pi  {\Upsilon}\left(\mho_{\cal {R}}\right),
\end{equation}
where $\alpha_{z}$ represents the total angle of jumps, set as $\pi$, and as $S_{rf}\rightarrow\infty$, the geodesic curvature $\mathfrak{h}\left(C_{\cal R}\right)$ can be expressed as the magnitude of the gradient of $\dot{C}_{\cal R}$ with respect to $\dot{C}_{\cal R}$ itself: $\left|\nabla_{\dot{C}_{\mathcal{R}}} \dot{C}_{\mathcal{R}}\right|$. The radial component of geodesic curvature can then be obtained as \cite{isUpadhyay:2023yhk}:

\begin{equation}\label{is44}
\left(\nabla_{\dot{C}_{\cal R}} \dot{C}_{\cal R}\right)^{r}=\dot{C}_{\cal R}^{\phi} \partial_{\varphi} \dot{C}_{\cal R}^{r}+\Gamma_{\phi \phi}^{r_{*}}\left(\dot{C}_{\cal R}^{\phi}\right)^{2}.
\end{equation}

For large ${\cal R}$, where $C_{\cal R}:=r(\phi)={\cal R}$ is constant, one can get $\left(\dot{C}_{R}^{\phi}\right)^{2}=\mathcal{F}^{-2}(r_{*})$. After making some algebra, the geodesic curvature can then be obtained as follows \cite{isGibbons:2008rj}:
\begin{equation}\label{is45}
\left(\nabla_{\dot{C}_{\cal R}^{r}} \dot{C}_{{\cal R}}^{r}\right)^{r} \rightarrow \frac{1}{{\cal R}},
\end{equation}
which means that $\mathfrak{h}\left(C_{\mathcal{R}}\right) \rightarrow \frac{1}{\mathcal{R}}$. Utilizing the optical metric \eqref{is4}, we get $d t={\cal R} d \phi$ leading to the following expression:

\begin{equation}\label{is46}
\mathfrak{h}(C_{\cal R})dt=\lim_{{\cal R}\to\infty}[\mathfrak{h}(C_{\cal R})dt]
         =\lim_{{\cal R}\to\infty}\left[\sqrt{\frac{ \tilde{g}^{\phi\phi}}{4\tilde{g}_{r_{*}r_{*}} }}\left(\frac{\partial\tilde{g}_{\phi\phi}}{\partial r_{*}}\right)\right]d\phi
         =d\phi.
\end{equation}

Taking into account all of the earlier findings, the GBth results in
\begin{equation}\label{is47}
\iint_{\mho_{{\cal R}}} \mathcal{K} d S_{rf}+\oint_{\partial \mho_{{\cal R}}} \mathfrak{h} d t\stackrel{{\cal R}\rightarrow \infty} {=}  \iint_{S_{rf:\infty}} \mathcal{K} d S_{rf}+\int_{0}^{\pi+\tilde{\delta}} d \phi.
\end{equation}
\begin{figure*}
    \centering
{\includegraphics[width=7.75cm]{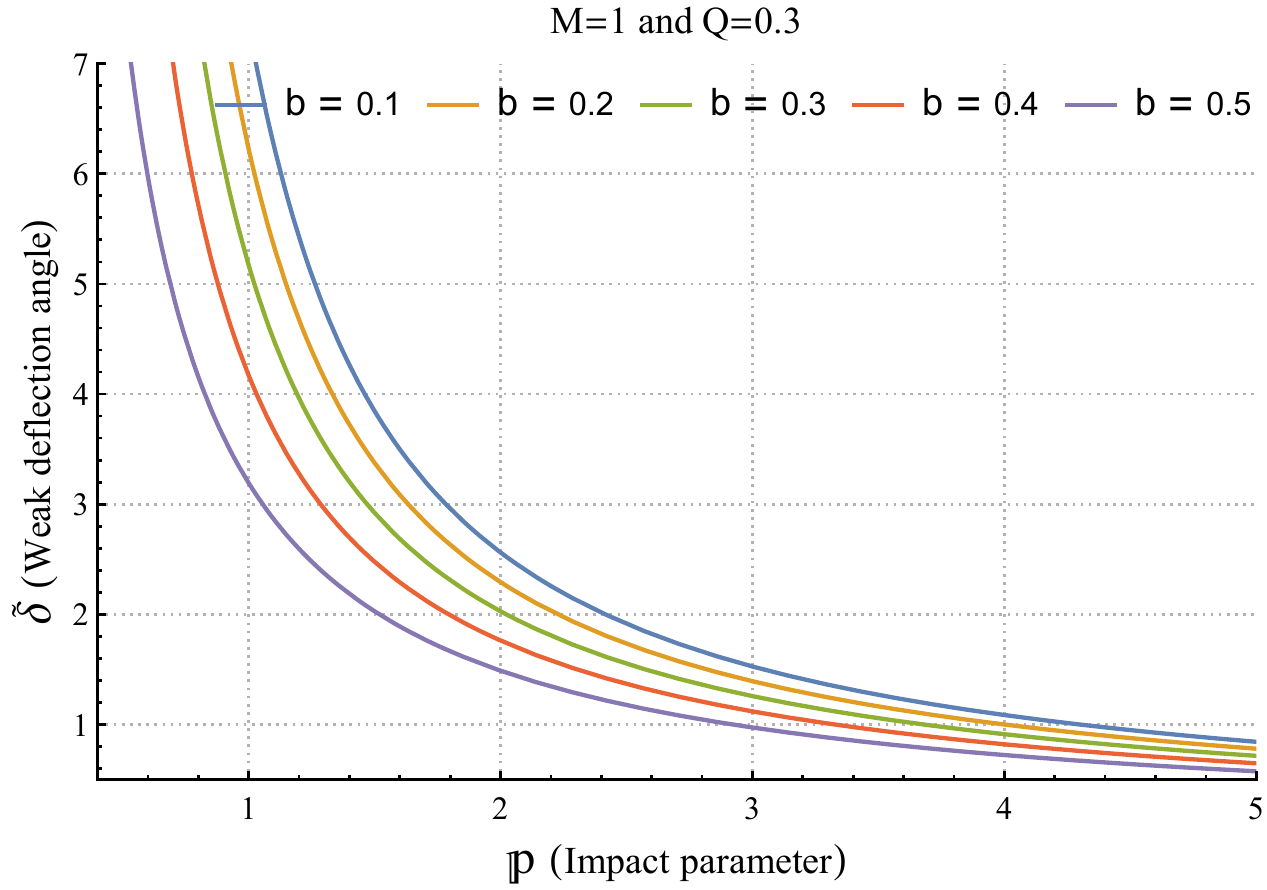} }\qquad
    {{\includegraphics[width=7.75cm]{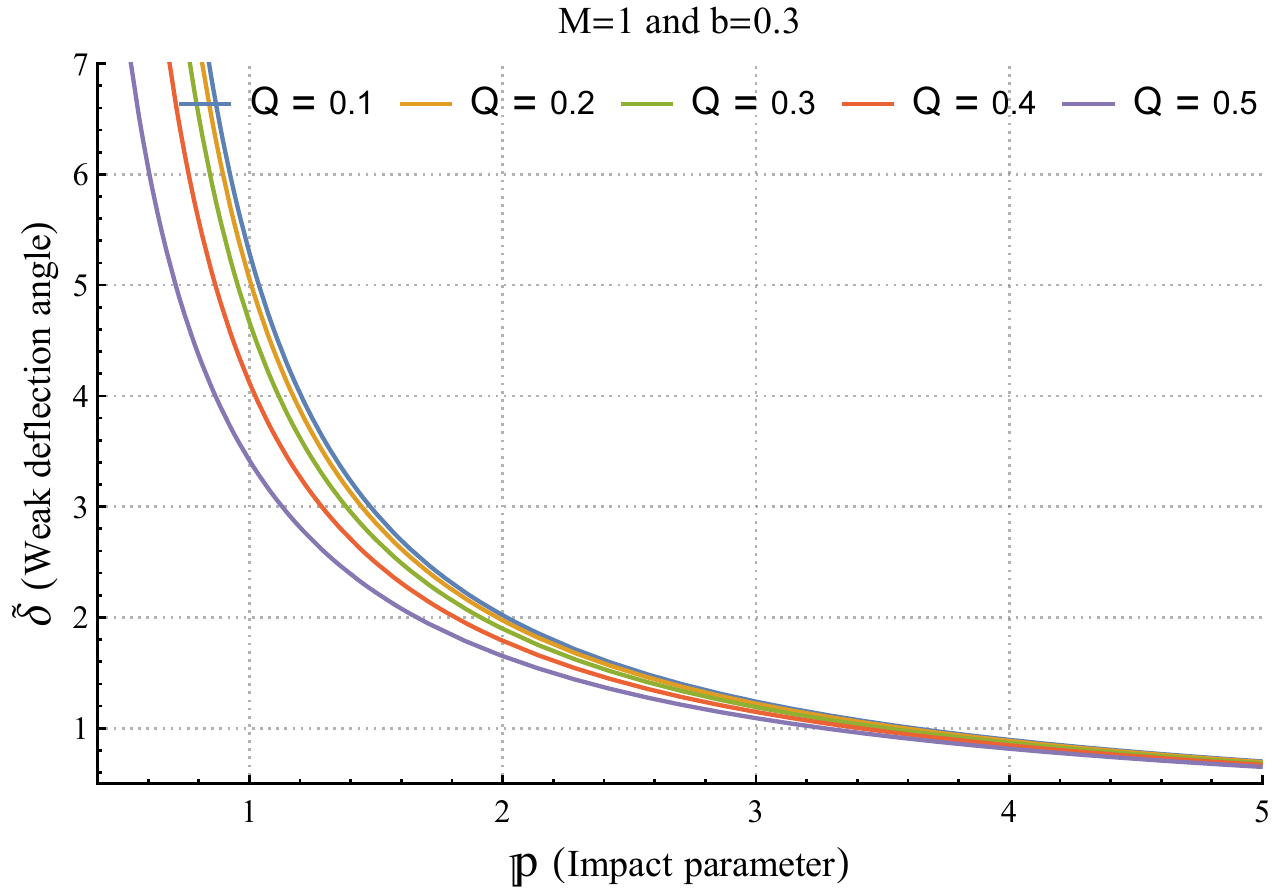}}}
    \caption{The profile of the deflection angle around the KN-like in gravity with a background KR field as a function of impact parameter $\mathfrak{p}$ for different values of Lorentz-violating parameter $b$ for fixed $Q=0.3$ (left panel) and of BH charge $Q$ for fixed $b=0.3$ (right panel). }
    \label{fig:def_an}
\end{figure*}

In the weak deflection limit scenario, the trajectory of a light ray is simplified to a straight line, represented by the equation $r(t)\equiv\mathfrak{B}=\frac{\mathfrak{p}}{\sin\varphi}$, where $\mathfrak{p}$ denotes the impact parameter \cite{isGibbons:2008rj,isOvgun:2020gjz}. Utilizing this expression, one can compute the deflection angle $\tilde{\delta}$ as follows \cite{isMandal:2023eae}:

\begin{equation}\label{is48}
\tilde{\delta}=-\int_{0}^{\pi} \int_{\mathfrak{B}}^{\infty} \mathcal{K} dS_{rf}=-\int_{0}^{\pi} \int_{\mathfrak{B}}^{\infty}\frac{\mathcal{K} \sqrt{\operatorname{det} \tilde{g}}}{F(r)} dr d \phi =-\int_{0}^{\pi} \int_{\mathfrak{B}}^{\infty}  \frac{r\mathcal{K}}{F(r)^\frac{3}{2}} d r d \phi.
\end{equation}
Therefore, in a non-plasma medium, for the spacetime of a spherically symmetric statically charged BH within a gravitational field coupled with a background KR field, the deflection angle \big(by using Eqs. \eqref{is0} and \eqref{is42} in Eq. \eqref{is48}\big) reads
\begin{equation}\label{is48}
\tilde{\delta}=-\int_{0}^{\pi} \int_{\mathfrak{B}}^{\infty}  \frac{1}{F(r)^\frac{3}{2}}\left(-2\frac{M\alpha}{{r}^{2}}+ 3\frac{{M}^{2}}{{r}^{3}} - 6\frac{{Q}^{2}{\alpha}^{2}M}{{r}^{4}} + 3\frac{{Q}^{2}{\alpha}^{3}}{{r}^{3}} + 2\frac{{Q}^{4}{\alpha}^{4}}{{r}^{5}}
\right)drd\phi,
\end{equation}
which can be approximated to the following form:
\begin{equation}\label{is49}
\tilde{\delta}\approx-\int_{0}^{\pi} \int_{\mathfrak{B}}^{\infty} \left(-\frac{2 M}{\sqrt{\alpha} r^2}-\frac{3 M^2}{\alpha^{(3 / 2)} r^3}+\frac{3 \alpha^{(3 / 2)} Q^2}{r^3}+\frac{6 \sqrt{\alpha} M Q^2}{r^4}-\frac{6 M^3}{\alpha^{(5 / 2)} r^4}\right)drd\phi.
\end{equation}
After making straightforward calculations, one can obtain the following deflection angle:
\begin{equation} \label{is50}
\tilde{\delta}\approx \frac{4 M}{\sqrt{\alpha} \mathfrak{p}}-\frac{3 \alpha^{\frac{3}{2}} \pi Q^2}{4 \mathfrak{p}^2}+\frac{3 \pi M^2}{4 \mathfrak{p}^2 \alpha^{\frac{3}{2}}}-\frac{8 \sqrt{\alpha} M Q^2}{3 \mathfrak{p}^3}+\frac{8 M^3}{3 \mathfrak{p}^3 \alpha^{\frac{5}{2}}}.
\end{equation}

Thus, one can immediately observe how the Lorentz-violating effects $\alpha$, and thus $b$, originating from the KR fields influence the deflection angle (\ref{is50}), which reduces to the Reissner-Nordstr\"{o}m BH case \cite{isJaved:2023iih,isJusufi:2015laa,isPang:2018jpm} in the absence of the $b = 0$. At this point, it is important to highlight that the GBth method can be applied to any asymptotically flat Riemannian optical metric due to its distinctive topological characteristics. 

We then turn to analyze the remarkable effect of Lorentz-violating parameter $b$ with BH charge on the weak deflection angle by adapting the developed method considered here. To this end we show the dependence of deflection angle profile from the impact parameter in Fig~\ref{fig:def_an}. For being more informative, in Fig~\ref{fig:def_an}, the left panel reflects the
impact of Lorentz-violating parameter $b$ with fixed $Q$ on the profile of deflection angle, while the right panel reflects the
impact of BH charge for fixed $b$. From Fig.~\ref{fig:def_an}, Lorentz-violating parameter $b$ affects the deflection angle so that it decreases significantly with increasing $b$. Note that $b$ and $Q$ have a similar physical effect thus resulting in decreasing the deflection angle. This behaviour can be related to the physical interpretation of Lorentz-violating parameter as repulsive gravitational charge.

\section{Conclusion} \label{sec6}
In this paper, we investigated the geodesic structure and deflection angle of the electrically charged BH \eqref{m1} in the presence of a nonzero vacuum expectation value background of the KR field. Various aspects were explored, including null geodesics, timelike geodesics, and the motion of charged particles in the BH close vicinity. Several key findings emerged from this investigation.

The analysis of null geodesics revealed intriguing insights into the photon sphere radius and the behavior of photons around charged BHs. It was demonstrated that the Lorentz-violating parameter $b$ significantly influenced the photon sphere radius, affecting the paths of photons in the gravitational field of the BH. Moreover, the investigation into timelike geodesics uncovered the nature of stable circular orbits for charged particles. The study revealed that the minimum radius for a stable circular orbit decreased with increasing Lorentz-violating parameter and charge, highlighting the sensitivity of these orbits to changes in the background field.

Based on the analysis, we showed that the Lorentz-violating parameter can alter the particle geodesics thus affecting the radius of ISCO, usually referred to one of observable properties. It does therefore could mimic the effects of BH spin parameter up to $a/M \sim 1$, thus having the same orbits as the one around a rotating Kerr BH. From observational perspectives, the current precision of measurements of highly spinning astrophysical BH candidates may not rule out the effect of Lorentz-violating parameter due to the fact that distant observers would not able to distinguish between a rotating Kerr BH from a static BH having the same spin in gravity with a background KR field. Our qualitative theoretical findings can help to explain astrophysical observations for distinguishing a variety of BH alternatives and to make useful astrophysical predictions. 

We further examined the motion of charged particles and analyzed the energy extracted by collisions near the BH horizon. The findings suggested that the Lorentz-violating parameter $b$ played a crucial role in determining the center-of-mass energy of colliding particles, indicating a decrease in energy as the Lorentz-violating parameter increased. Furthermore, the study provided valuable insights into the behavior of charged particles in the vicinity of those BHs under the influence of both gravitational and electromagnetic fields. By investigating the trajectories of charged particles, our research shed light on the dynamics of charged particle motion and the influence of the background KR field on their trajectories. We also studied gravitational lensing phenomenon for this BH in the weak field approximation by utilizing the GBth. We analytically showed how the Lorentz-violating effects $\alpha$, and thus $b$, originating from the KR fields influence the deflection angle (\ref{is50}). \textcolor{blue}{Our results showed that both photon orbits and the deflection angle decrease significantly with an increasing Lorentz-violating parameter acting as a repulsive gravitational charge. }

The research presented in this paper contributed to our understanding of the intricate interplay between gravitational and electromagnetic fields in the vicinity of charged BHs with a background KR field. The findings advanced our knowledge of fundamental physics in extreme gravitational environments. \textcolor{blue}{These theoretical findings may also help constrain the validity of alternative models of spacetime geometry for BHs in explaining quantum effects, non-linear interactions and astrophysical observations, contributing to the field and adding a unique perspective to the existing literature.} Moving forward, potential future directions for this study could involve investigating the effects of other background fields on the geodesic structure and particle dynamics of charged BHs. Additionally, exploring the implications of these findings for observational astronomy and astrophysical phenomena, such as accretion disks  \cite{isAbramowicz:2011xu,isOrtega-Rodriguez:2006kdi} and quasinormal modes \cite{aa11,isBerti:2009kk,aa12,isSakalli:2022xrb,isKokkotas:1999bd,aa13}, could further enhance our understanding of the behavior of matter and radiation in the vicinity of such BHs. Moreover, extending the analysis to incorporate quantum effects and non-linear interactions, like greybody radiation \cite{isMaldacena:1996ix,isAl-Badawi:2024iax,isUniyal:2022xnq,isAl-Badawi:2024kdw}, could offer new perspectives on the behavior of charged BHs in realistic astrophysical scenarios. Those tasks are slated for our near-future agenda.

\begin{acknowledgments}
We are thankful to the Editor and anonymous Referees for their constructive suggestions and comments.
%
The research is supported by the National Natural Science Foundation of China under Grant No. 11675143 and the National Key Research and Development Program of China under Grant No. 2020YFC2201503. \.{I}.~S. would like to acknowledge networking support of COST Actions CA21106 and CA22113. He also thanks to T\"{U}B\.{I}TAK, ANKOS, and SCOAP3 for their support.
\end{acknowledgments}

\end{document}